\documentclass[12pt]{article}
\usepackage{graphicx}
\usepackage{cite}
\usepackage{color}
\newcommand{\xrm}[1]{{\textstyle \mbox{\rm #1}}}

\newcommand{\bm}[1]{\mbox{\boldmath $#1$}}

\newcommand{\zt}{\tilde{Z}(57)}
\begin{document}
\title{Substructures from weak interactions\\
in light of possible threshold signals at LEP and LHC}
\author{
Eef van Beveren\\
{\normalsize\it Centro de F\'{\i}sica Computacional}\\
{\normalsize\it Departamento de F\'{\i}sica, Universidade de Coimbra}\\
{\normalsize\it P-3004-516 Coimbra, Portugal}\\
{\small eef@teor.fis.uc.pt}\\ [.3cm]
Susana Coito and George Rupp\\
{\normalsize\it Centro de F\'{\i}sica das Interac\c{c}\~{o}es Fundamentais}\\
{\normalsize\it Instituto Superior T\'{e}cnico,
Universidade T\'{e}cnica de Lisboa}\\
{\normalsize\it Edif\'{\i}cio Ci\^{e}ncia,
P-1049-001 Lisboa, Portugal}\\
{\small susana.coito@ist.utl.pt, george@ist.utl.pt}\\ [.3cm]
}

\maketitle

\begin{abstract}
We present indications of possible substructures from weak interactions,
by inspecting LEP and LHC data and inferring threshold effects due to
the production of pairs of composite heavy gauge bosons $W^{\pm}$,
$Z$ and their hypothetical partners with different spin.
Thus, we find possible evidence of scalar or pseudoscalar partners
of the $W^{\pm}$ and the $Z$, viz. at 53 and 57 GeV, respectively.
Additionally, data may indicate excited states of the $Z$ at 210
and 240 GeV.
\end{abstract}

\section{Introduction}
\label{intro}

A theoretical model of threshold enhancements in hadronic production
amplitudes, based on quark-antiquark pair creation, was formulated in
Ref.~\cite{AP323p1215} and further developed in
Refs.~\cite{EPL81p61002,EPL84p51002}. The model shows that one must
expect non-resonant enhancements in the amplitudes just above
pair-creation thresholds.
Experimental evidence of this phenomenon is scarce though,
since it needs event counts with high statistics and good resolution.
Nevertheless, in some cases signals, albeit feeble, can be seen in
experimental data.

\subsection{The reaction \bm{e^{+}e^{-}\to D\bar{D}}}

\begin{figure}[htbp]
\begin{center}
\begin{tabular}{c}
\scalebox{0.8}{\includegraphics{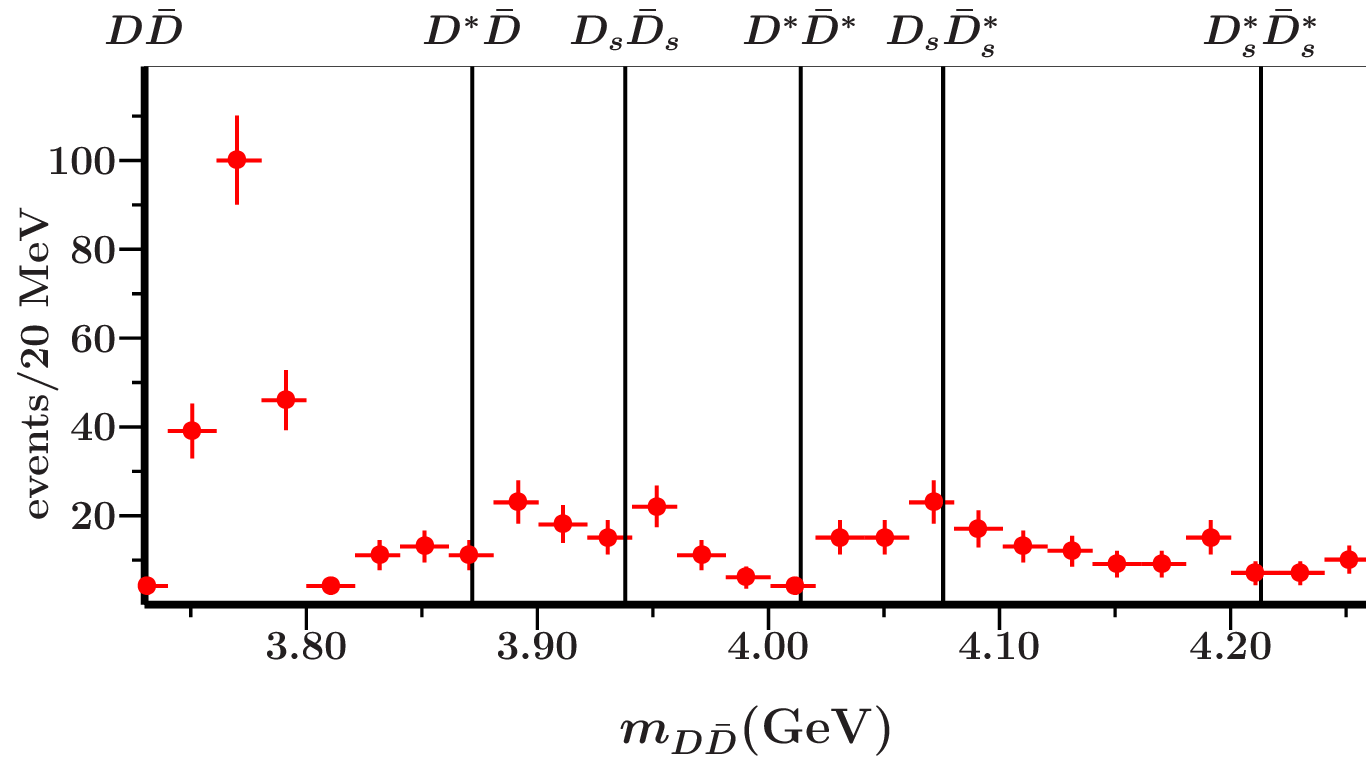}}
\end{tabular}
\end{center}
\caption[]{\small
Experimental data for $D\bar{D}$ production in ISR
obtained by the BABAR Collaboration \cite{PRD79p092001}.
The open-charm thresholds, indicated by vertical lines,
come at respectively 3.73 ($D\bar{D}$),
3.87 ($D^{\ast}\bar{D}$),
3.94 ($D_{s}\bar{D}_{s}$),
4.02 ($D^{\ast}\bar{D}^{\ast}$),
4.08 ($D^{\ast}_{s}\bar{D}_{s}$)
and
4.21 ($D^{\ast}_{s}\bar{D}^{\ast}_{s}$) GeV.}
\label{kleinDD}
\end{figure}

The $e^{+}e^{-}$ annihilation into open-charm pairs can be observed
right above the $D\bar{D}$ threshold at 3.73 GeV.
We assume here that the reaction takes place via a photon
and the $c\bar{c}$ propagator,
through the creation of a light $q\bar{q}$ pair.
However, as has been discussed previously \cite{ZPC21p291},
many competing configurations may be formed,
increasing in number for higher invariant masses and so
reducing the intensity of the specific decay into $D\bar{D}$,
contrary to what would be expected from available phase space.
Furthermore, experiment seems to indicate that stable open-charm
hadrons have a higher probability to be formed near threshold,
i.e., when kinetic energy is almost zero.
Hence, were it not for phase space and the centrifugal barrier,
$D\bar{D}$ pairs would be produced most likely just above threshold.
In Fig.~\ref{kleinDD} we depict the experimental observation
published by the BABAR Collaboration in Ref.~\cite{PRD79p092001}.
We see a strong enhancement peaking at about 3.77 GeV.
However, this peak is usually considered to be due to a $c\bar{c}$ resonance,
viz.\ the $\psi (3770)$, and not a threshold enhancement.
At higher energies, just above the $D^{\ast}\bar{D}$, $D_{s}\bar{D}_{s}$,
and $D^{\ast}\bar{D}^{\ast}$ thresholds, data also suggest enhancements,
but clearly much higher statistics and resolution are necessary to confirm
such structures.

Furthermore, one can observe a sharp dip at about 3.81 GeV, and also
that the amplitude is not enhanced right above the $D^{\ast}_{s}\bar{D}_{s}$
and $D^{\ast}_{s}\bar{D}^{\ast}_{s}$ thresholds, at 4.08~GeV and 4.21 GeV,
respectively.
The latter observations might be explained in a similar fashion, namely
through interference effects between nearby resonances and threshold
enhancements. At least, this appears to be the case for the sharp dip at
3.81 GeV, as confirmed by the BES Collaboration \cite{ARXIV08070494}.
\begin{figure}[htbp]
\begin{center}
\begin{tabular}{c}
\scalebox{0.8}{\includegraphics{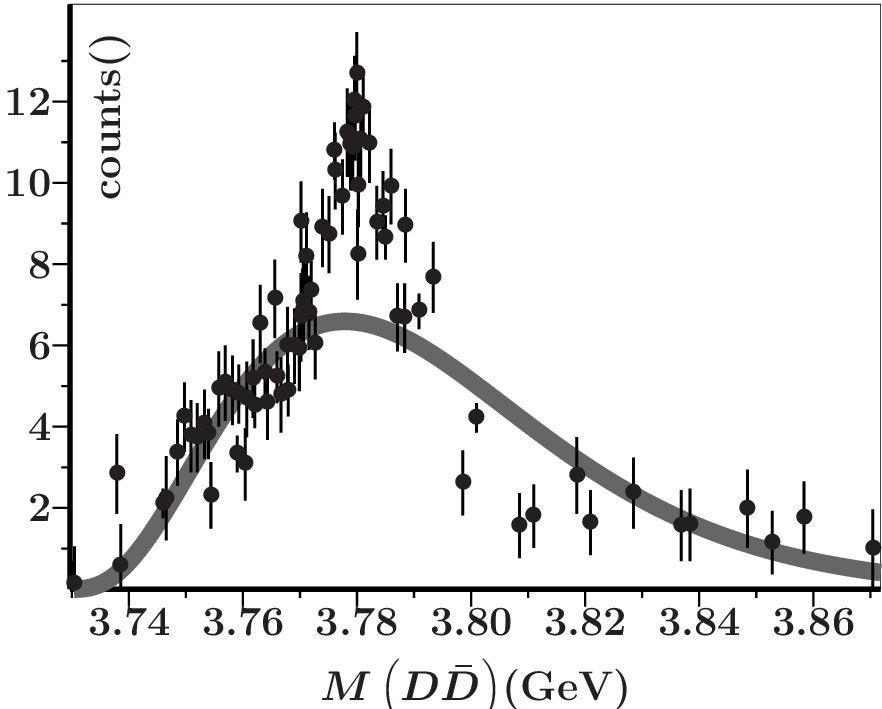}}
\end{tabular}
\end{center}
\caption[]{\small
Experimental data after background subtraction,
obtained by the BES Collaboration \cite{ARXIV08070494}.
The Breit-Wigner (BW) parameters of the $c\bar{c}$ resonance are
$M\left(\psi (1D)\right) =3.781$ GeV and
$\Gamma\left(\psi (1D)\right) =17$ MeV.
}
\label{bes}
\end{figure}

To our knowledge, BES was the first to observe
that the $\psi (3770)$ cross section is built up
by two different amplitudes, viz.\ a relatively broad signal and a
rather narrow $c\bar{c}$ resonance.
For the narrow resonance, which probably corresponds to
the well-established $\psi (1D)(3770)$,
BES measured a central resonance position
of $3781.0\pm 1.3\pm 0.5$ MeV
and a width of $19.3\pm 3.1\pm 0.1$ MeV (their solution 2).
If the latter parameters are indeed confirmed,
it will be yet another observation
of a quark-antiquark resonance width
that is very different from the world average
($83.9\pm 2.4$ MeV \cite{PLB667p1} in this case),
after a similar result was obtained by the BABAR Collaboration
\cite{PRL102p012001}, for $b\bar{b}$ resonances.
Concerning the broader structure, BES indicated for
their solution no.~2 a central resonance position of
$3762.6\pm 11.8\pm 0.5$ MeV and a width of $49.9\pm 32.1\pm 0.1$ MeV.
The signal significance for the new enhancement is $7.6\sigma$
(solution 2). On the theoretical side, a diresonance \cite{PRD78p116014}
or heavy molecular state \cite{EPJC63p411} have been proposed for
this new structure.

Moreover, in the latter BES paper \cite{ARXIV08070494}, the existence of
conflicting results with respect to the branching fraction
for non-$D\bar{D}$ hadronic decays of the $\psi (1D)(3770)$ was emphasised.
On the one hand, the total branching fraction
for exclusive non-$D\bar{D}$ modes has been measured
to be less than 2\% \cite{PLB605p63,PRD74p012005}.
But on the other hand, for inclusive non-$D\bar{D}$ decay modes,
values of about 15\% have been found \cite{PRD76p122002,PLB659p74}.
According to BES, this apparent discrepancy may be partly
due to the assumption that the line shape above the $D\bar{D}$ threshold
is the result of one simple resonance.
In Fig.~\ref{bes} we show our interpretation
\cite{PRD80p074001} of the BES data,
namely a resonance, the $\psi (3770)$,
on top of the enhancement above the $D\bar{D}$ threshold.
The interference between the two signals of very different origin
is well visible for energies in the interval 3.79--3.82 GeV.

Also note that, in spite of being extracted from different reactions,
the data displayed in Figs.~\ref{kleinDD} and \ref{bes}
agree in their shape on various details, viz.\
a dip at about 3.81 GeV as well as the relative heights
of the signals at 3.77--3.78 GeV and 3.85 GeV.
Nevertheless, without the existence of the data collected in Fig.~\ref{bes},
an interpretation in terms of interfering threshold and resonance enhancements
for the data shown in Fig.~\ref{kleinDD}
would hardly be plausible. We stress once more that data
with better resolution and higher statistics are indispensible
for any serious theoretical analysis.

\subsection{The reaction \bm{e^{+}e^{-}\to B\bar{B}}}

\begin{figure}[htbp]
\begin{center}
\begin{tabular}{c}
\scalebox{0.6}{\includegraphics{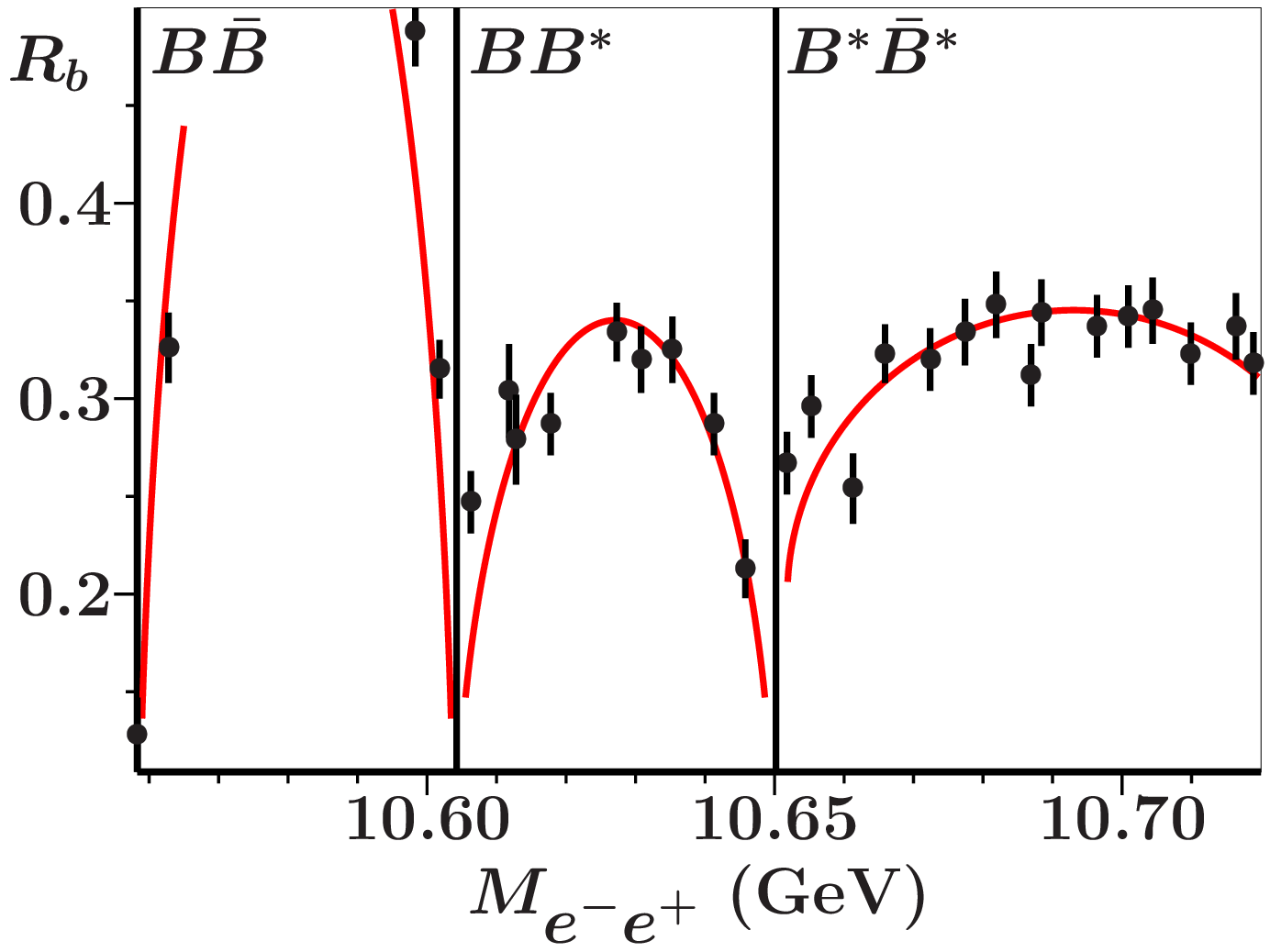}}\\ [-15pt]
\end{tabular}
\end{center}
\caption[]{\small
Experimental data for the process $e^{+}e^{-}\to b\bar{b}$
measured by the BABAR Collaboration \cite{PRL102p012001}.
The vertical lines indicate the $BB^{\ast}$ and $B^{\ast}B^{\ast}$
thresholds, as indicated in the figure.
The eye-guiding lines reflect our interpretation
of the data, and do not represent fits.}
\label{morebabar}
\end{figure}

A similar phenomenon can be observed in Fig.~\ref{morebabar},
in which we show data for the process $e^{+}e^{-}\to b\bar{b}$,
measured and analysed by the BABAR Collaboration \cite{PRL102p012001}.
As also remarked in their paper,
the large statistics and the small energy steps of the
scan make it possible to clearly observe the two dips
at the opening of the thresholds corresponding
to the $B\bar{B}^{\ast}+\bar{B}B^{\ast}$
and $B^{\ast}\bar{B}^{\ast}$ channels.

In the figure, we have cut off the huge peak at 10.58 GeV,
in order to concentrate better on
the details of the other two enhancements, viz.\
at 10.63 GeV and 10.69 GeV, respectively.
Near the $BB^{\ast}$ threshold, we thus observe that
the $B\bar{B}$ signal rapidly
decreases for increasing invariant mass,
whereas the $BB^{\ast}$ signal behaves the opposite way, i.e.,
it grows fast for increasing invariant mass
just above threshold.
At the $B^{\ast}B^{\ast}$ threshold, this phenomenon repeats itself,
now with respect to the $BB^{\ast}$ signal.

So we conclude that $B\bar{B}$ production in $e^{+}e^{-}$ annihilation
can mainly be observed within the invariant-mass window
delimited by the $B\bar{B}$ and $BB^{\ast}$ thresholds.
Similarly, $BB^{\ast}$ production has an equally wide window formed
by the $BB^{\ast}$ and $B^{\ast}B^{\ast}$ thresholds.
The window for $B^{\ast}B^{\ast}$ production is somewhat wider,
since the next threshold concerns the $B_{s}B_{s}$ channel,
which lies considerably higher. Therefore, the enhancement peaking at
about 10.69~GeV is broader than the ones at 10.58 and 10.63~GeV.
\begin{figure}[htbp]
\begin{center}
\begin{tabular}{c}
\scalebox{0.6}{\includegraphics{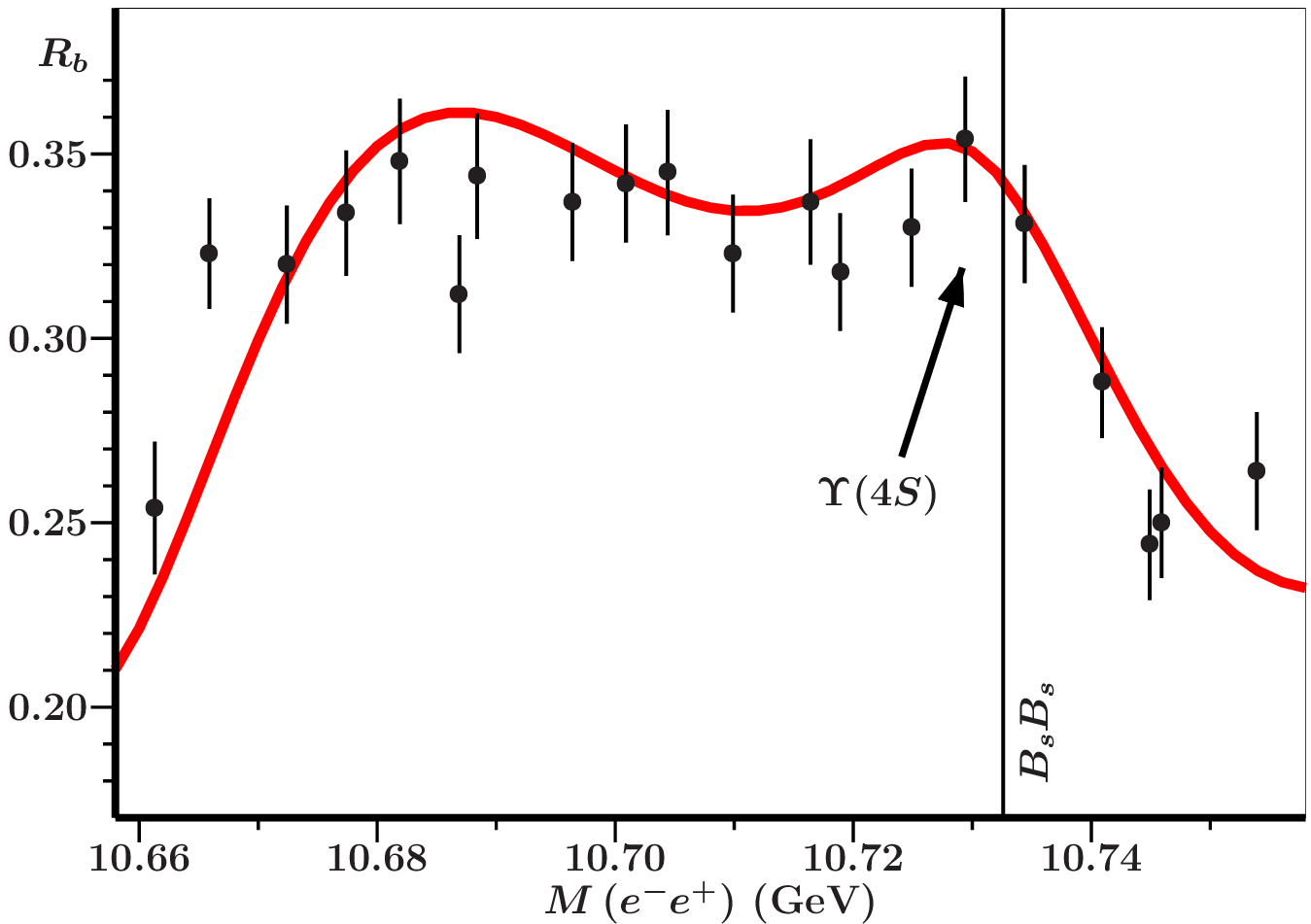}}
\end{tabular}
\end{center}
\caption[]{\small
Detail of the data ($\bullet$)
for hadron production in electron-positron annihilation
published by the BABAR Collaboration \cite{PRL102p012001}.
The solid line is part of a global fit to the $R_{b}$ data,
discussed in Ref.~\cite{ARXIV09100967}.
}
\label{elephit}
\end{figure}

With the insight we have gained from analysing BABAR's bottomonium data,
we now revisit the $D\bar{D}$ data in Fig.~\ref{kleinDD}.
Indeed we see that, for the $D^{\ast}\bar{D}$ threshold at 3.87 GeV,
the $D_{s}\bar{D}_{s}$ threshold at 3.94 GeV, and
the $D^{\ast}\bar{D}^{\ast}$ threshold at 4.02 GeV,
the data might have a similar interpretation as those in Fig.~\ref{morebabar},
although a much higher accuracy is needed to confirm this.
However, the data right above the $D^{\ast}_{s}\bar{D}_{s}$ threshold at
4.08 GeV behave quite differently, which we study in the following.

In Fig.~\ref{elephit} we show the BABAR $e^-e^+$ data for energies
around the $B_{s}\bar{B}_{s}$ threshold. We observe that at and just
above threshold the data behave very similarly to those
displayed in Fig.~\ref{bes} concerning the $\psi(3770)$ resonance.
In particular, in both sets of data there is a clear dip
just above the local peak in the data. We are convinced that these
patterns stem from interference, that is, in the bottomonium case
the $\Upsilon (4S)$ resonance interferes with
the expected $B_{s}\bar{B}_{s}$ threshold threshold enhancement. We thus
conclude that the enhancement at 10.58 GeV is not due to the
$\Upsilon (4S)$, but rather amounts to a non-resonant, apparent peak
resulting from the openings of the $B\bar{B}$ and $BB^\ast$ thresholds.
We also believe that a corresponding interference pattern between
the $\psi (4040)$ $c\bar{c}$ resonance and the expected line shape
of the $D^{\ast}_{s}\bar{D}_{s}$ threshold enhancement at 4.08 GeV
is what one observes in Fig.~\ref{kleinDD}, but much better data are
needed to study this in detail.

\subsection{Charmonium resonances}
\label{Charm}

The enhancements at thresholds for pairs of mesons
which are stable with respect to strong interactions
are clearly visible in the data.
However, when the pair consists of one or two unstable mesons
with considerable decay widths, the signal is less pronounced.
Thus, the charmonium spectrum, in particular its lower part,
contains enhancements having different origins.
Now, in principle we do not expect a lot of correlation
between the $c\bar{c}$ resonance masses and threshold openings.
Consequently, the situation depicted in Fig.~\ref{bes},
where the $\psi (3770)$ $c\bar{c}$ resonance comes out
on top of the $D\bar{D}$ threshold enhancement, appears to be accidental.
\begin{figure}[htbp]
\begin{center}
\begin{tabular}{c}
\scalebox{1.0}{\includegraphics{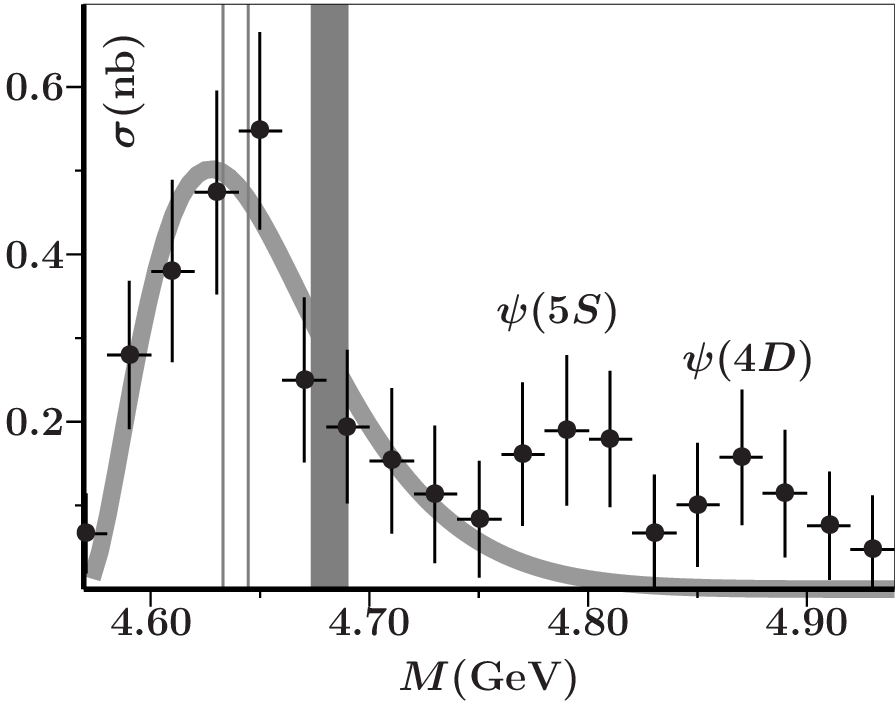}}\\ [-15pt]
\end{tabular}
\end{center}
\caption[]{\small
Experimental cross sections for the reaction
$e^{+}e^{-}\to\Lambda_{c}^{+}\Lambda_{c}^{-}$,
obtained by the Belle Collaboration \cite{PRL101p172001},
The non-resonant threshold enhancement is relatively large,
whereas the $\psi (5S)$ and $\psi (4D)$ resonances are more modest.
Besides the opening of the $\Lambda_{c}^{+}\Lambda_{c}^{-}$ threshold
at 4.573 GeV, one must also consider here the thresholds for
$D^{\ast}_{s0}(2317)^{+}\bar{D}^{\ast}_{s0}(2317)^{-}$ at 4.635 GeV,
$D^{\ast}_{s}D_{s1}(2536)$ at 4.647 GeV,
and $D^{\ast}_{s}D^{\ast}_{s2}(2573)$ at 4.684 GeV.
The latter threshold has some spreading
because of the 17 MeV width of the $D^{\ast}_{s2}(2573)$.
}
\label{LcLc}
\end{figure}

In Ref.~\cite{PRD80p074001} we discussed the production
of the $\psi (5S)$ and $\psi (4D)$ resonances
in the $e^{+}e^{-}\to\Lambda_{c}^{+}\Lambda_{c}^{-}$ cross
section reported by the Belle Collaboration \cite{PRL101p172001}.
This case is interesting because it demonstrates
the point that threshold enhancements, in this case
the $\Lambda_{c}^{+}\Lambda_{c}^{-}$ threshold at 4.573 GeV,
result in a signal that is considerably larger than the resonance peaks.
Furthermore, also several two-meson channels open
in the relevant mass interval depicted in Fig.~\ref{LcLc},
namely $D^{\ast}_{s0}(2317)^{+}\bar{D}^{\ast}_{s0}(2317)^{-}$ at 4.635 GeV,
$D^{\ast}_{s}D_{s1}(2536)$ at 4.647 GeV,
and $D^{\ast}_{s}D^{\ast}_{s2}(2573)$ at 4.684 GeV,
besides channels with very unstable final-state mesons.
So in view of our previous discussion
we expect much more structure than what we actually observe in the data
of Fig.~\ref{LcLc}. Therefore, much better resolution and statistics
are probably required that than obtained by Belle,
before one can draw conclusions on the possible existence of $c\bar{c}$
resonances at or near the $\Lambda_{c}^{+}\Lambda_{c}^{-}$ threshold.

\section{Threshold enhancements at LEP and LHC}

Now that we have acquired the taste of interpreting hadronic data in a
way very different from common practice in data analysis,  we may search
for similar effects in the gauge-boson and Higgs sectors of the Standard
Model (SM), as a possible consequence of compositeness. Thus, we shall look
at LEP and LHC data, with the purpose of convincing the reader that a
5--7 $\sigma$ enhancement at about 125 GeV does not necessarily imply
the discovery of a particle or resonance with that mass.
To put it differently, only more detailed data may allow one to
distinguish between the various scenarios proposed in the literature,
not only around 125 GeV, but at any invariant mass accessible by LHC,
and of course in many different channels.

More than a decade ago, the ALEPH and L3 Collaborations reported
an excess of data in the reaction $e^{+}e^{-}\to Z^{\ast}\to HZ$,
consistent with the production
of Higgs bosons with a mass of about 114 GeV
\cite{PLB495p1,PLB495p18}.
The data were collected using, respectively,
the ALEPH and L3 detectors at LEP
in the year 2000 at centre-of-mass energies
which range from 200 GeV to 209 GeV.
Here, we elaborate on the idea that the signal at 114 GeV
could be the onset of a threshold enhancement,
corresponding to the creation of a pair of spin-zero bosons
with a mass of about 57 GeV.
Composite heavy gauge bosons and their spin-zero partners,
the latter with a mass in the range 50--60 GeV,
have been considered long ago \cite{PLB135p313}
and studied in numerous works (see e.g.\ Refs.\
\cite{PLB141p455,PRAMANA23p607,PRD36p969,NCA90p49,PRD39p3458,PRL57p3245}).
To date, no experimental evidence of their existence has been reported.
\clearpage

\subsection{The reaction \bm{e^{+}e^{-}\to b\bar{b}} at LEP}
\label{Rb}

\begin{figure}[htbp]
\begin{center}
\begin{tabular}{c}
\scalebox{0.8}{\includegraphics{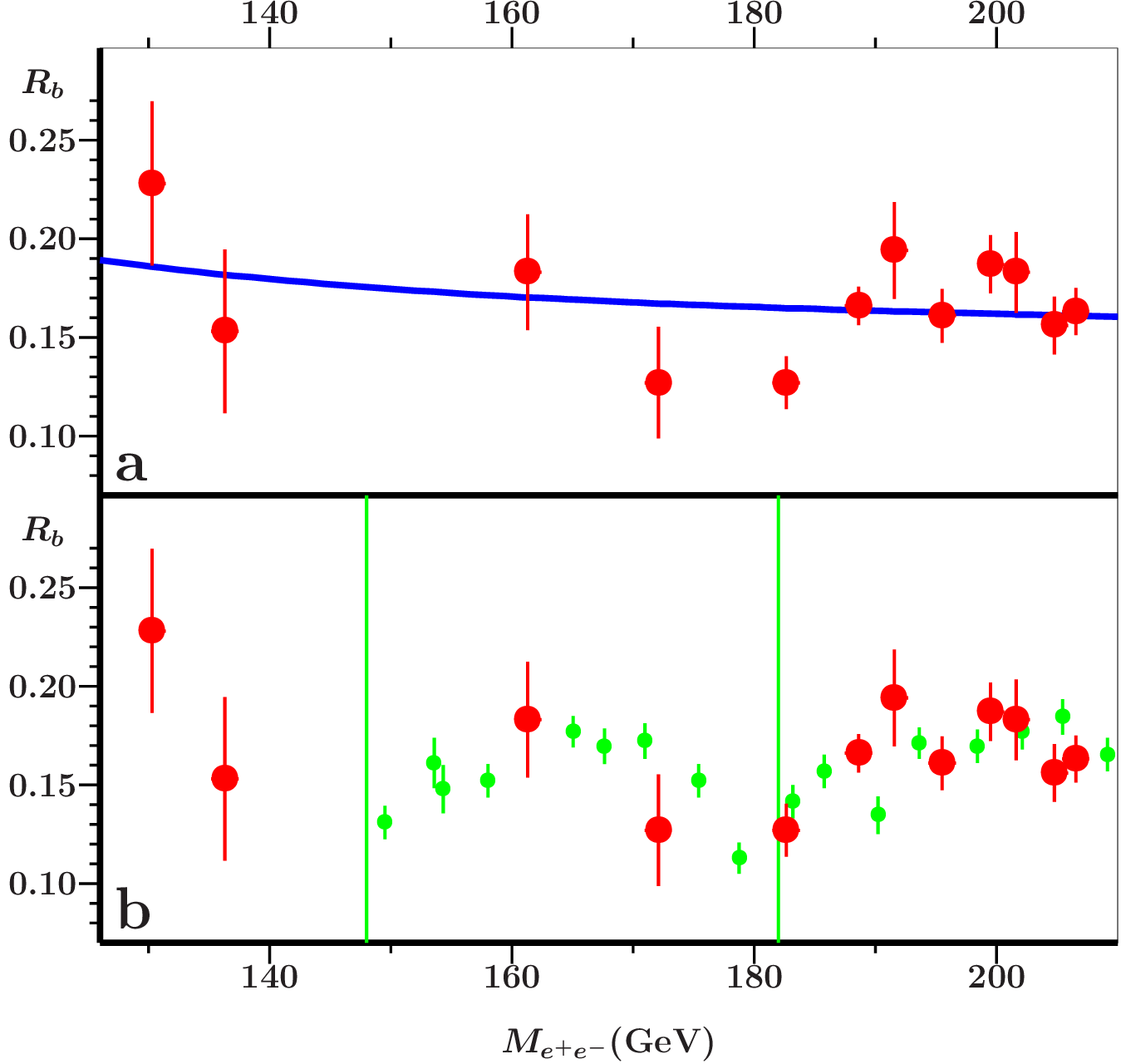}}\\ [-15pt]
\end{tabular}
\end{center}
\caption[]{\small
Experimental data for $b\bar{b}$ production in $e^{+}e^{-}$
obtained by the DELPHI Collaboration at LEP \cite{EPJC60p1}
(large dots, red in colour version).
In a) the SM prediction is represented by the solid curve.
In b) we additionally display the $R_{b}$ data (small grey dots, green
in colour version) of the BABAR Collaboration
(see Fig.~\ref{morebabar}), for which the invariant $e^{+}e^{-}$ masses
have been scaled up such that the $BB^{\ast}$ and $B^{\ast}B^{\ast}$
thresholds now lie at 148 and 182 GeV, respectively,
and the values of $R_{b}$ have been scaled down so
as to coincide with the DELPHI data.
The threshold masses
are indicated by vertical lines.
}
\label{RbDelphi}
\end{figure}

In Ref.~\cite{EPJC60p1}, the DELPHI Collaboration
published LEP data for $b\bar{b}$ production in $e^{+}e^{-}$.
These data, as well as the SM prediction
from the same DELPHI paper,
are collected in Fig.~\ref{RbDelphi}a.

The same data are also collected in Fig.~\ref{RbDelphi}b.
However, in addition we include here, as a kind of physically
motivated ``Monte Carlo'', the $R_{b}$ data of the BABAR Collaboration
\cite{PRL102p012001}, after having scaled up the invariant $e^{+}e^{-}$
masses such that the $BB^{\ast}$ and $B^{\ast}B^{\ast}$ thresholds
come out at 148 and 182 GeV, respectively,
and also scaled down the values of $R_{b}$ so
as to coincide with the DELPHI data.
We had already depicted the former
data in Fig.~\ref{morebabar}, in accordance with our interpretation
in terms of threshold enhancements.
Now, the quality of the DELPHI $R_{b}$ data certainly does not allow
to distinguish between the rather flat SM prediction and
our threshold description involving additional bosons. In the latter
scenario, there would exist a light partner of the $Z$ boson with a mass
of about 57 GeV,  a ``$\zt$'', so as to account
for a $Z\zt$ threshold at about 148 GeV.
Evidently, the threshold at 182 GeV corresponds to $ZZ$ production.
%\clearpage

\subsection{Diphoton events at LHC analysed by CMS and ATLAS}
\label{diphoton}

\begin{figure}[htbp]
\begin{center}
\begin{tabular}{c}
\scalebox{0.8}{\includegraphics{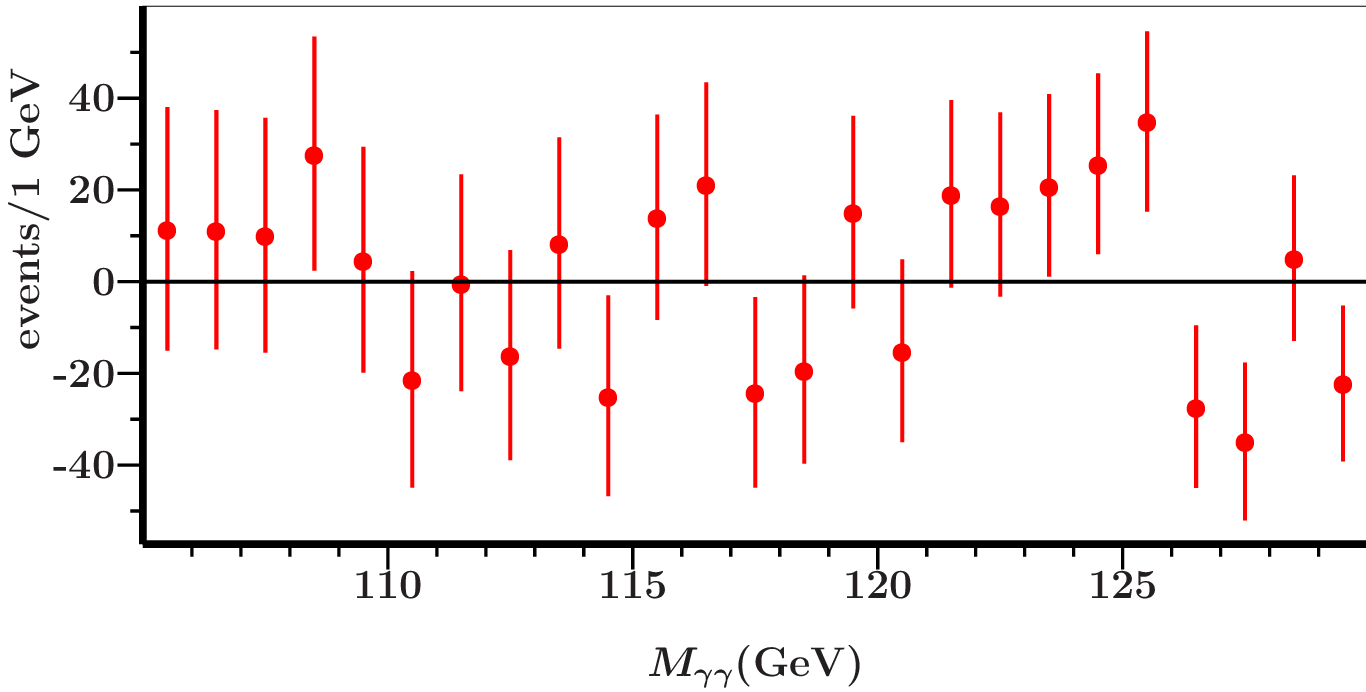}}\\ [-15pt]
\end{tabular}
\end{center}
\caption[]{\small
Experimental data for diphoton production in $pp$,
obtained by the CMS Collaboration at LHC \cite{PLB710p403}
}
\label{CMSgg}
\end{figure}

In Ref.~\cite{PLB710p403} the CMS Collaboration collected diphoton events
corresponding to an integrated luminosity of 4.8 fb$^{-1}$.
Diphoton triggers with asymmetric transverse energy $E_T$,
thresholds, and complementary photon selections were used.
Furthermore, CMS did a polynomial fit through their data.
In Fig.~\ref{CMSgg} we show the resulting event distribution,
after subtraction of the fit.
Now, in the first place we observe that the signal does not have
enough statistics for any firm observation of possible structures in the
data.  Nevertheless, the signal appears to suggest a dip somewhere between
110 and 115 GeV, and also an enhancement with maximum at about 125--126 GeV.
The latter enhancement is interpreted by CMS as a signal of a Higgs-like
boson.  However, we will consider it a threshold enhancement,
with threshold opening around 114 GeV.  But either interpretation of the
signal shown in Fig.~\ref{CMSgg} will need a lot more statistics
and higher resolution to be unequivocally confirmed.

\begin{figure}[htbp]
\begin{center}
\begin{tabular}{c}
\scalebox{0.8}{\includegraphics{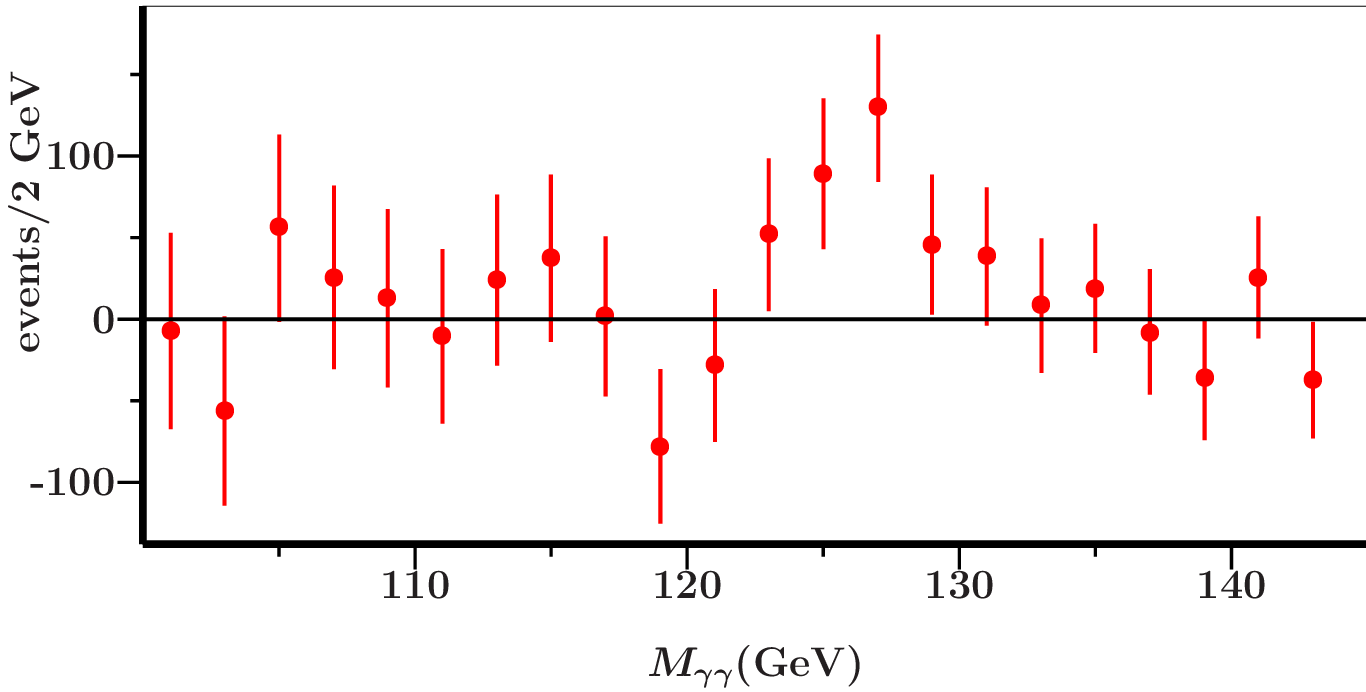}}\\ [-15pt]
\end{tabular}
\end{center}
\caption[]{\small
Experimental data for diphoton production in $pp$,
obtained by the ATLAS Collaboration at LHC \cite{PLB716p1}
}
\label{ATLASgg}
\end{figure}
A similar analysis, performed by the ATLAS Collaboration
\cite{PLB716p1}, yields the result depicted in Fig.~\ref{ATLASgg}.
Here, with slightly better statistics, though lower resolution,
we find that the dip comes out closer to 120 GeV, while the
enhancement also peaks at about 125--126 GeV. Again, definite
conclusions may only be drawn when much improved data will become
available.  Note that the onset of an enhancement at 114 GeV corresponds
to a threshold at 2$\times$57 GeV.

\subsection{The reaction \bm{e^{+}e^{-}\to\tau^{+}\tau^{-}} at LEP}

\begin{figure}[htbp]
\begin{center}
\begin{tabular}{c}
\scalebox{0.8}{\includegraphics{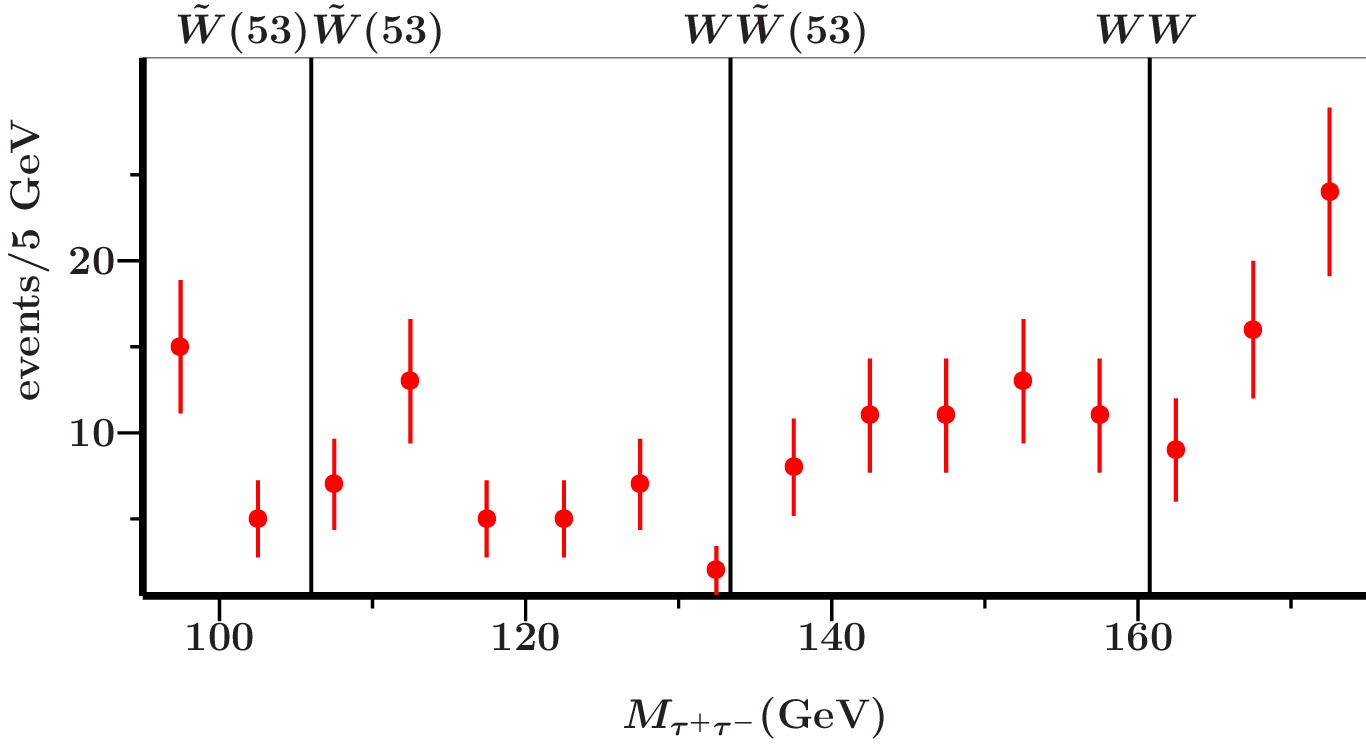}}\\ [-15pt]
\end{tabular}
\end{center}
\caption[]{\small
Experimental data for $\tau^{+}\tau^{-}$ production in $e^{+}e^{-}$,
obtained by the L3 Collaboration at LEP \cite{PLB479p101}
(large dots, red in colour version).
}
\label{tautauL3}
\end{figure}

In Ref.~\cite{PLB479p101} the L3 Collaboration
published experimental data, obtained at LEP,
for $\tau^{+}\tau^{-}$ production in $e^{+}e^{-}$.
We collect these data in Fig.~\ref{tautauL3},
in which we also indicate possible thresholds at about 106, 133,
and 161 GeV, so values quite different from those proposed in Sect.~\ref{Rb}.
Nevertheless, we may recognise the energy of 161 GeV as the
$W^{+}W^{-}$ production threshold. Then, the other two energies may
correspond to thresholds involving one or two lighter $\tilde{W}^\pm(53)$
bosons, $\tilde{W}^\pm(53)$ being a hypothetical (pseudo)scalar partner
of the $W^{\pm}$, compatible with compositeness.
Again, we must insist on the necessity to produce much better data
in order to conclude whether a weak substructure has been discovered,
or just a Higgs-like boson.
%\clearpage

\subsection{The four-lepton signal at LHC}

\begin{figure}[htbp]
\begin{center}
\begin{tabular}{c}
\scalebox{0.8}{\includegraphics{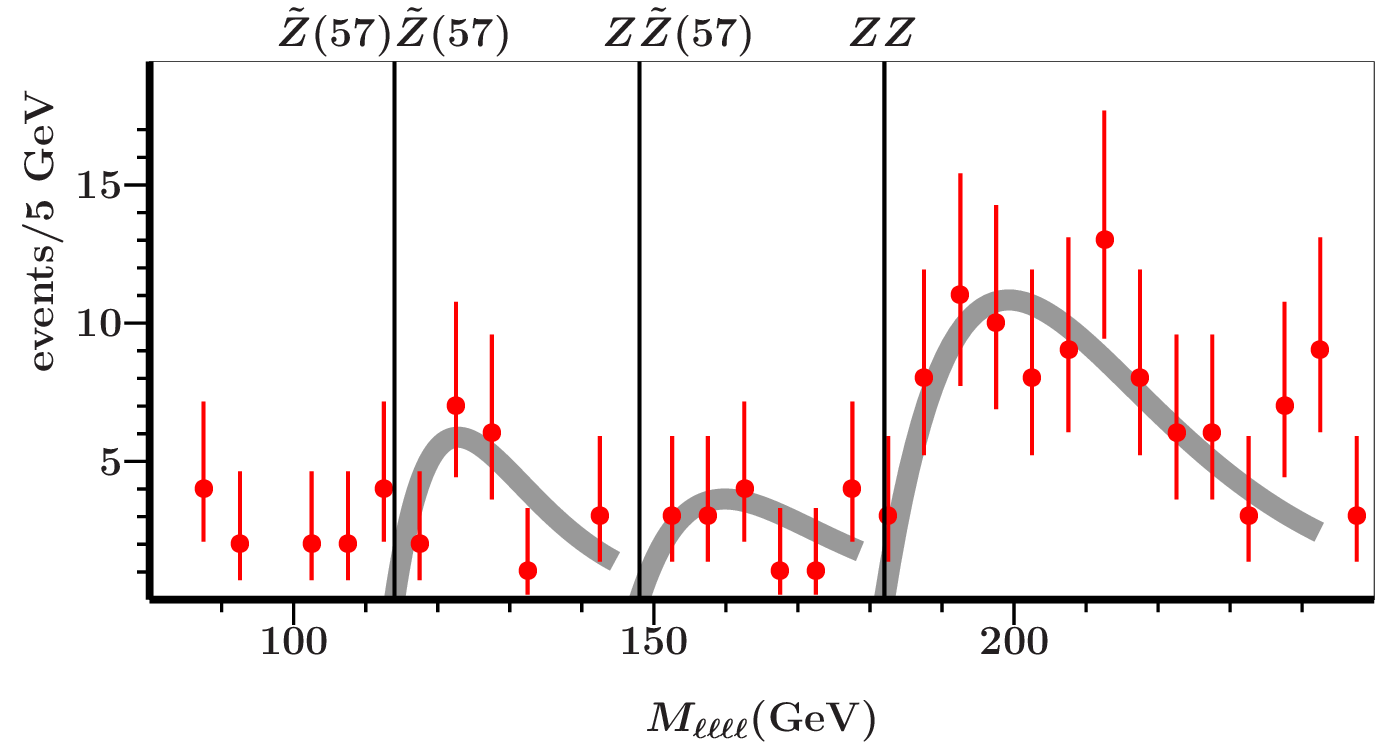}}\\ [-15pt]
\end{tabular}
\end{center}
\caption[]{\small
Experimental data for four-lepton production in $pp$,
obtained by the ATLAS Collaboration \cite{PLB716p1}.
We have indicated by vertical lines the thresholds
of respectively, from left to right, $\zt\zt$,
$\zt Z$, and $ZZ$.}
\label{ATLASZZbar}
\end{figure}
\begin{figure}[htbp]
\begin{center}
\begin{tabular}{c}
\scalebox{1.0}{\includegraphics{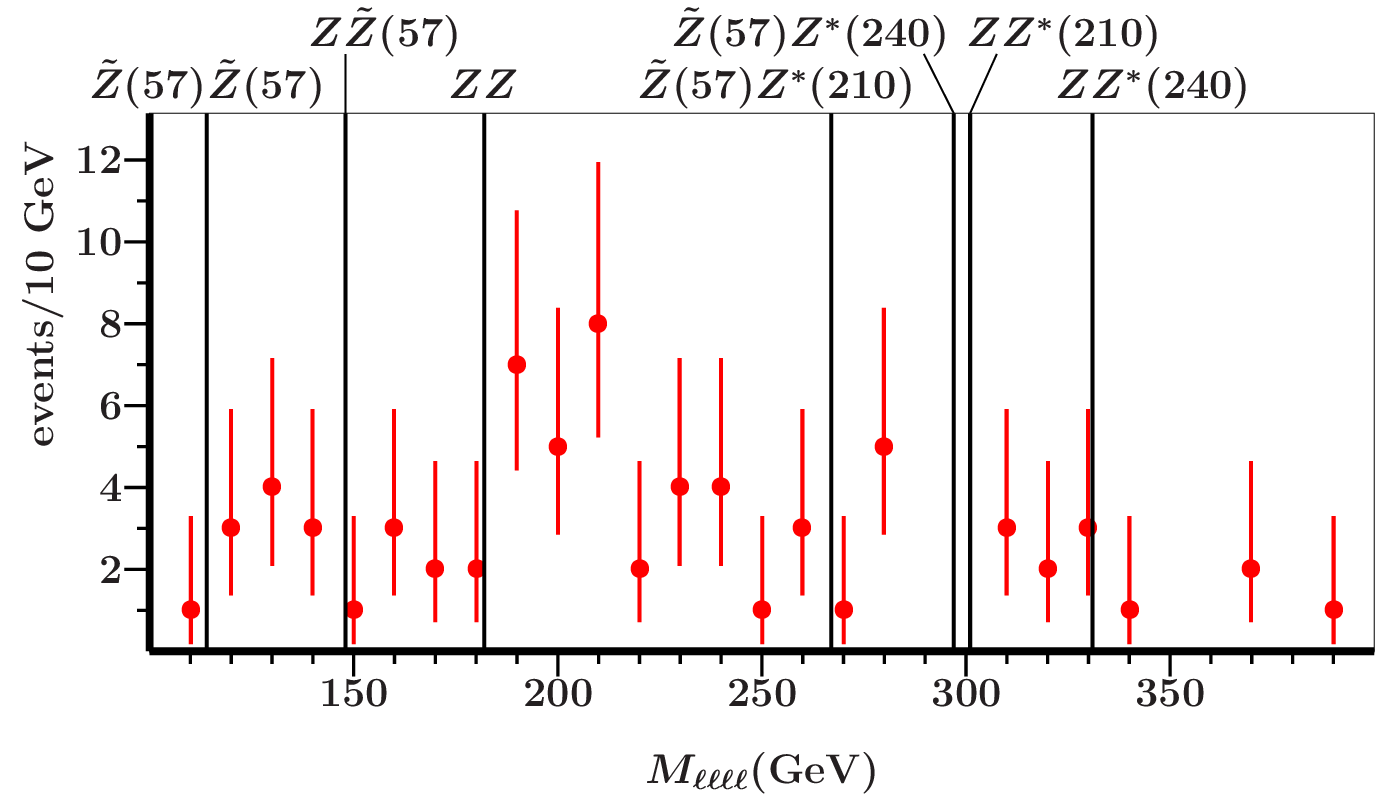}}\\ [-15pt]
\end{tabular}
\end{center}
\caption[]{\small
Experimental data for four-lepton production in $pp$,
obtained by the CMS Collaboration \cite{ARXIV12052907}.
We have indicated by vertical lines the thresholds
of, from left to right, $\zt\zt$,
$\zt Z$, $ZZ$, $\zt Z^\ast(210)$, $\zt Z^\ast(240)$,
$ZZ^\ast(210)$ and $ZZ^\ast(240)$.}
\label{CMSZZbar}
\end{figure}

The ATLAS \cite{PLB716p1} and CMS \cite{ARXIV12052907} Collaborations
have also published experimental data on four-lepton production in $pp$
collisions, by selecting events with two pairs of isolated leptons, each
of which comprising two leptons with the same flavour (either $e$ or $\mu$)
and opposite charge.
The same-flavour and opposite-charge lepton pair
with invariant mass closest to the $Z$-boson mass of about 91 GeV,
is required to have an invariant mass between 50 GeV and 106 GeV.
The invariant mass of the remaining same-flavour
opposite-charge lepton pair must not exceed 115 GeV
in the event selection of candidates.
In Fig.~\ref{ATLASZZbar} and Fig.~\ref{CMSZZbar} we depict
the resulting four-lepton invariant mass distributions.

From Fig.~\ref{ATLASZZbar} one vaguely recognises threshold dips
at 114, 148, and 182 GeV, for the modes $\zt\zt$,
$\zt Z$, and $ZZ$, respectively.
Hence, this is at least not inconsistent with the result
of Sect.~\ref{Rb}.
However, it is also clear that highly improved statistics
and resolution are needed to confirm this scenario.
Nevertheless, in analogy with the result of Sect.~\ref{Charm}
and in particular Fig.~\ref{LcLc},
we also seem to observe two $Z$-like excitations
in the ATLAS data for four-lepton production in $pp$ collisions
(Fig.~\ref{ATLASZZbar}), namely at about 210 GeV and 240 GeV,
being excited states of the $\zt$ and the $Z$, respectively.

The existence of the latter resonances would imply
further thresholds at 267, 297, 301, and 331 GeV.
These thresholds cannot be observed in Fig.~\ref{ATLASZZbar}, which
only shows results up to 250 GeV.
However, the CMS data, albeit with even lower statistics,
concern four-lepton production up to 400 GeV,
which we depict in Fig.~\ref{CMSZZbar}, together with
the production thresholds for $\zt\zt$,
$\zt Z$, $ZZ$, $\zt Z^\ast(210)$, $\zt Z^\ast(240)$,
$ZZ^\ast(210)$, and $ZZ^\ast(240)$.
One can observe that there is a striking agreement
between the dips in the data and the threshold openings,
a lot more data would be needed to confirm such a pattern.

Note that the enhancements at 210 and 240 GeV are also visible
in the CMS data (see Fig.~\ref{CMSZZbar}), albeit with lower statistics.

\section{Summary and conclusions}

In the foregoing, we have argued on the basis of many experimental data
that threshold effects may be responsible for several structures in both
strong- and weak-interaction processes, which therefore can be
of a non-resonant nature. However, the quality of the data is generally
too poor to unmistakably confirm such an interpretation, which underlines
the need for new measurements, with higher statistics and better resolution.

In the mesonic sector, most mainstream models do not contemplate such
threshold effects, and so resort to exotic configurations whenever
a newly observed meson does not fit in with standard spectroscopic
predictions. However, we  have shown in numerous published and unpublished
works that practically all mesonic enhancements can be either explained as
unitarised quark-antiquark states, mass shifted in the complex energy plane
due to open and closed two-meson decay channels, or as non-resonant peaks
from threshold openings and inelasticity/depletion effects (see
Ref.~\cite{ARXIV10112360} for a classification of four different types of
enhancements, and also further references). Moreover, in all these works we
have successfully employed a constant radial spacing for the bare
quark-antiquark states, resulting from a flavour-independent oscillator
frequency of about 190 MeV, as well as a transition radius for quark-pair
creation in two-meson decay characterised by a scalar quantum of about 38 MeV
\cite{ARXIV10095191,ARXIV11021863,ARXIV12021739}.

The main purpose of the present paper was to show that the described threshold
phenomena, observed for mesons, may also have an application to heavy
gauge bosons at high energies, where the weak interactions are in fact not
weak anymore. Confirmation of such effects would indicate compositeness in this
sector of the SM, too. Moreover, it could offer an explanation for the
enhancement around 125 GeV seen by ATLAS and CMS at LHC, as a possible
alternative to a generally accepted Higgs-like particle of such a mass.
A consequence of this scenario would be the existence of gauge-boson partners,
of lower mass and with different quantum numbers, being either scalars or
pseudoscalars. The LHC and LEP data we have presented above indeed hint at the
existence of such partners, namely a $\tilde{W}(53)$ at 53 GeV and
a $\tilde{Z}(57)$ at 57 GeV. Moreover, there are also indications, albeit
feeble, of $Z$-like recurrencies, viz.\ at about 210 and 240 GeV. But as we
have said several times, only much improved statistics and resolution
can settle this issue of a possible weak substructure and its manifestation
through new heavy bosons.

Apart from assuming that the possible partners of the $Z$ and $W$ are bosons,
we have not made any attempt to discuss the nature of the corresponding
compositeness, as we find it premature in view of the insufficient
statistics and resolution of the data.
However, if a $Z^\ast(240)$ exists, it seems natural that this excitation
of the $Z$ boson will show up in the dilepton data as clearly as 
the $Z$ itself. Such data may be available from the huge amount of dileptons
produced at LHC. Hence, the possible existence of a boson at 240 GeV can
probably be determined from the already measured data.

\section{Acknowledgements}

We are grateful for the precise measurements
and data analyses of the L3, ALEPH, DELPHI, BES, Belle,
BABAR, CMS, and ATLAS Collaborations,
which made the present analysis possible.
This work was supported in part by the {\it Funda\c{c}\~{a}o para a
Ci\^{e}ncia e a Tecnologia} \/of the {\it Minist\'{e}rio da Ci\^{e}ncia,
Tecnologia e Ensino Superior} \/of Portugal, under contract
CERN/\-FP/\-123576/\-2011.

\newcommand{\pubprt}[4]{#1 {\bf #2}, #3 (#4)}
\newcommand{\ertbid}[4]{[Erratum-ibid.~#1 {\bf #2}, #3 (#4)]}
\def\AP{Ann.\ Phys.}
\def\EPJC{Eur.\ Phys.\ J.\ C}
\def\EPL{Europhys.\ Lett.}
\def\NCA{Nuovo Cim.\ A}
\def\PLB{Phys.\ Lett.\ B}
\def\PRAMANA{Pramana}
\def\PRD{Phys.\ Rev.\ D}
\def\PRL{Phys.\ Rev.\ Lett.}
\def\ZPC{Z.\ Phys.\ C}

\end{document}